\documentclass[a4paper,11pt]{article}
\usepackage{jinstpub} 
\usepackage{romannum}
\usepackage{lineno}

\title{\boldmath Design and Implementation of a UDP-Based Command Interface for the INO ICAL Experiment}

\author[a,1]{Yuvaraj Elangovan,\note{Corresponding author.}}
\author[a]{Mandar Saraf,}
\author[a]{B. Satyanarayana,} 
\author[a]{S.S. Upadhya,} 
\author[a]{Nagaraj Panyam,} 
\author[a]{Ravindra Shinde,} 
\author[a]{Gobinda Majumder,} 
\author[a]{D. Sil,} 
\author[a]{Pathaleswar,} 
\author[a]{K.C. Ravindran,} 
\author[a]{Upendra Gokhale,} 
\author[a]{and Pavan Kumar} 
\affiliation[a]{Tata Institute of Fundamental Research, Mumbai, India}

\emailAdd{yue8@pitt.edu}

\abstract{Efficient command interface is a critical requirement for experiments employing a large number of front-end DAQ modules and control servers. In the context of the INO ICAL (India-based Neutrino Observatory Iron Calorimeter) experiment, this involves 28,800 Resistive Plate Chamber (RPC), charged particle detectors. The acquisition and control of these detectors are facilitated through Front End data acquisition modules known as RPC-DAQs. These modules are equipped with Ethernet interfaces for data and command connectivity to a server. Each module acts as a network node with a unique IP address. The collective group of hundreds of modules is controlled by a common server over a Local Area Network (LAN).  UDP (User Datagram Protocol) is the most commonly used networking protocol which supports Multicast as well as Unicast and can be easily adapted to INO ICAL Experiment. A  server can send commands to a group of DAQs or any particular DAQ. But UDP may have the problem of packet loss and reliability. To mitigate these issues, this paper suggests a simple approach that modifies the UDP protocol by implementing a handshaking scheme and checksum, similar to those found in more reliable protocols like TCP. The proposed solution optimizes the use of UDP as a reliable command interface in the INO ICAL experiment, ensuring seamless data acquisition and control. Also, this paper shows the performance study of the custom hybrid UDP Command Interface in the prototype ICAL experiment called Mini Iron Calorimeter (mini-ICAL) which houses 20 units of RPCs and electronics. This work not only addresses the challenges of the INO ICAL experiment but also underscores the adaptability and robustness of the proposed protocol for usage in mini-ICAL and beyond.}


\begin{document}
\maketitle
\flushbottom

\section{Introduction}
\label{sec:intro}
\begin{figure}[htbp]
\centering
\includegraphics[width=.8\textwidth]{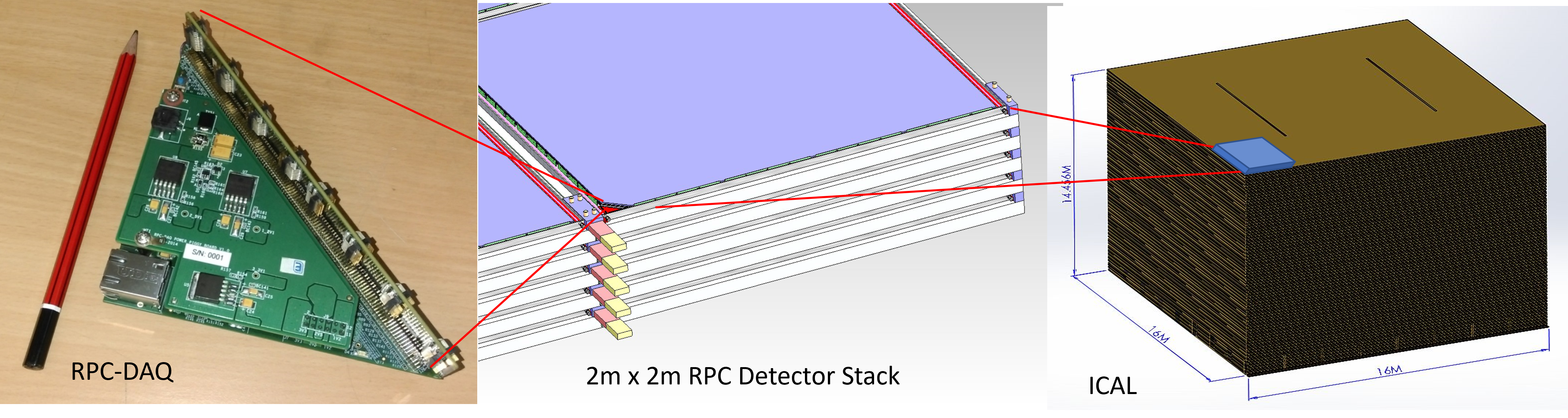}
\caption{RPC-DAQ position in INO ICAL experiment. \label{fig:1}}
\end{figure}

Data acquisition and control are the crucial components of the readout electronics in the particle physics experiments. Within the context of the INO ICAL (India-Based Neutrino Observatory Iron Calorimeter) experiment, a substantial number of sensors are employed known as RPCs (Resistive Plate Chambers). Each of these RPCs is interfaced with a digital front end known as RPC-DAQ~\cite{rpcdaq} (RPC Data Acquisition Module) as shown in Figure~\ref{fig:1}. The ICAL experiment uses a vast array of 28,800 individual RPCs with RPC-DAQ Front End, each serving as a detector unit.

These DAQ modules with wired Ethernet interface~\cite{ethernet} are positioned near the detectors they serve as shown in Figure~\ref{fig:1}. Each of these RPC-DAQs is configured with a unique IP address. This web of DAQ modules along with the interconnected detectors, collectively forms what is commonly referred to as the sensor network nodes. The primary objective of ICAL experiment is to track charged particles over a large area, so it is important to ensure the continuous and synchronized operation of these detectors. Central command servers are developed to oversee the group performance of RPC-DAQs modules. In principle these servers act as the  control and configuration nodes for the entire ICAL.
Standard TCP/IP networking protocols are readily available, but the demands of the experiment, factors such as timing precision, data reliability, and group management, require a careful and tailored approach. Due to this, it is clear that customizing standard protocols is the ideal approach to meet the needs of the ICAL experiment.

\section{Selection of Protocol and Customization}
\label{sec:proto}
In high-performance distributed data acquisition systems, the choice of communication protocol must balance reliability, resource usage, timing determinism, and group addressing. TCP is a well-established and reliable protocol stack offering in-order delivery, congestion control, and retransmission mechanisms. However, it is connection-oriented, resource-heavy and does not support native multicast, which limits its suitability in large embedded networks. In contrast, UDP is lightweight, stateless, and supports multicast, making it ideal for one-to-many control messaging. UDP provides an optional CRC-based checksum to verify data integrity, although its use is not mandatory. By combining the reliability features of TCP with the group managing features of UDP a hybrid protocol was designed called Hybrid Protocol based Command Interface (HPCI).
Several projects have previously explored the implementation of similar reliability mechanism over UDP to overcome the limitations of TCP in multicast or real-time systems. For example, the Plan9 operating system developed a lightweight Reliable UDP (RUDP) protocol~\cite{rudp_plan9,draft_rudp} that introduced acknowledgment and retransmission schemes to ensure data integrity while maintaining simplicity. Similarly, CERN's IPBus protocol~\cite{ipbus_cern} uses a UDP-based stack for slow-control operations in FPGA-based data acquisition systems. Other efforts such as those described in~\cite{larrea2015ipbus,jinst2015_udp,ieee2021_udpdaq,arxiv2020_neutron} demonstrate customized solutions in high-energy and nuclear physics contexts, where real-time reliability and synchronization are critical.

While these implementations address similar challenges our HPCI protocol is optimized for deployment on low-end FPGAs such as Intel Cyclone IV with minimal logic usage and memory overhead. The DAQ modules used in INO ICAL are equipped with limited resources and operate within tightly controlled timing budgets. Hence, our design emphasizes a deterministic handshake scheme, lightweight CRC-16 error detection, and dual-socket command acknowledgment mechanism without requiring a full TCP/IP stack. These constraints and the large scale of the deployment (over 28,000 DAQs) necessitate a specialized protocol that balances reliability, scalability, and hardware simplicity. This distinguishes HPCI from more generalized or software-centric solutions. HPCI is also used as a command-and-control server for other subsystems like Trigger system ~\cite{trigger} and Data Concentrator ~\cite{dc} to synchronise the data acquisition in the ICAL experiment.

\section{HPCI Model Overview}
\label{sec:conf}

\begin{figure}[htbp]
\centering
\includegraphics[width=.6\textwidth]{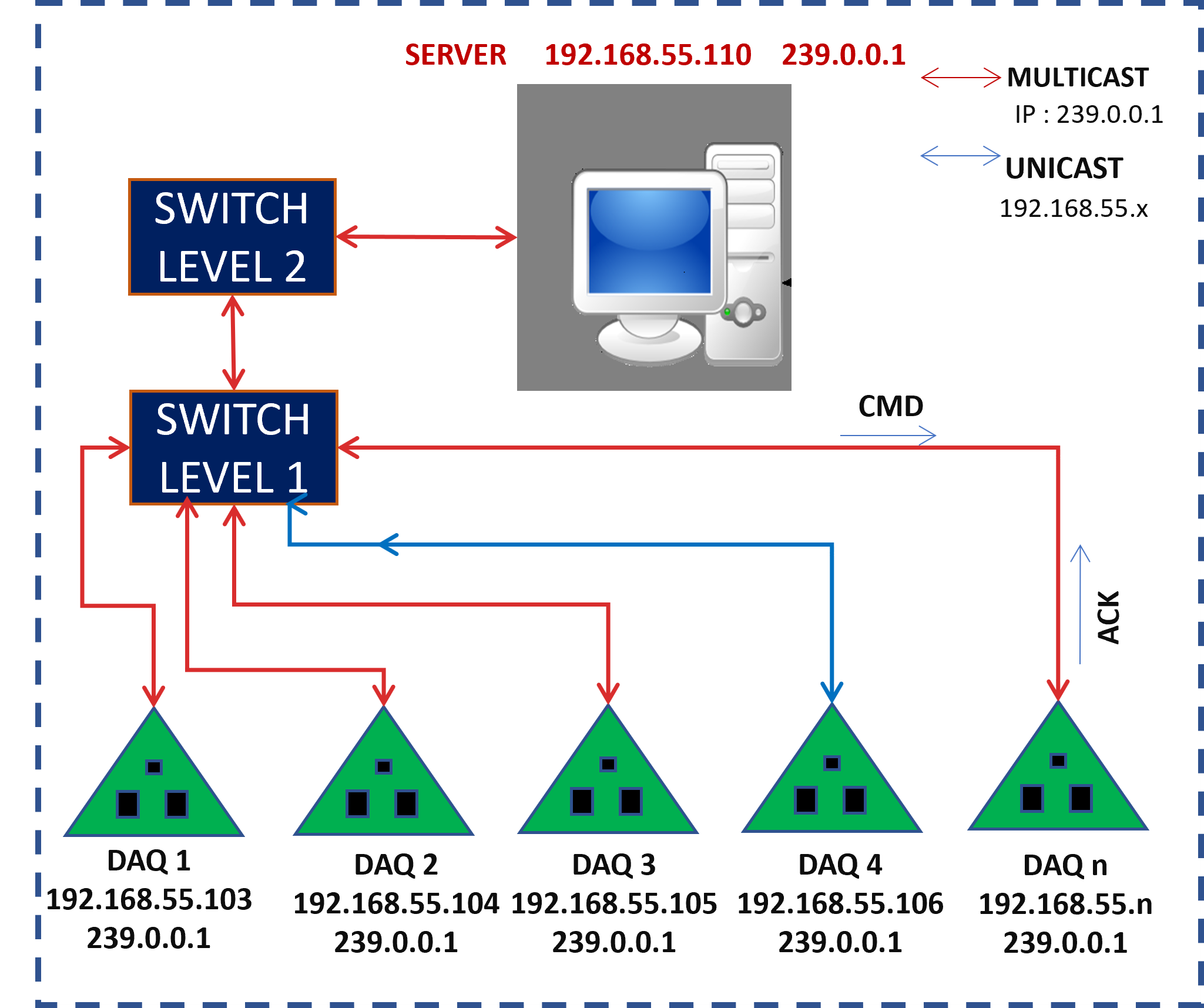}
\caption{Prototype HPCI Model with 5 RPC-DAQs.\label{fig:2}}
\end{figure}

In the HPCI model, multiple DAQs are interconnected with a central command server as shown in Figure~\ref{fig:2}.  Users have the capability to issue global commands from the server, which can be distributed to all the connected DAQs through Multicast grouping. Each RPC-DAQ module opens an UDP socket that is configured to receive Multicast packets. These sockets are programmed to join a shared Multicast group using Internet Group Management Protocol (IGMP), with a designated group IP address (e.g., 239.0.0.1). Joining this group enables the DAQs to receive Multicast commands sent from the central control server to the Multicast group address, thereby allowing simultaneous command delivery to all nodes.

Apart from Multicast the HPCI model supports transmission of commands to individual DAQs through Unicast mode. For example, initiating a "RUN START" command can be sent in Multicast mode, making sure that the same command packet reaches all DAQs simultaneously. In situations where a particular DAQ shows irregular behavior during a run, users can switch to Unicast mode to request the status of that specific DAQ. To support this feature, each DAQ must have two open sockets—one used to handle Multicast packets and the other handles Unicast packets. Acknowledgments of both types of commands are sent to the server through the Unicast socket.

The pilot HPCI model infrastructure was developed to handle 100 RPC-DAQs simultaneously. Ethernet switches are used at both Level 1 and Level 2, to handle Gigabit data transfer and communication efficiency. All the RPC-DAQs are configured with IPv4 addresses to support standard network protocols. Each DAQ has the feature of remotely programmable network configuration, allowing users to modify network configurations when needed. Network Configuration like IP Address, MAC Address, GATEWAY and SUBNET are permanently stored in the flash memory of each DAQ . During the boot sequence, the FPGA processor reads network setup from a dedicated location of its Flash memory. This helps in configuring correct network information every time while DAQ is booting. Dedicated commands are available to user to change the network configuration stored in the flash memory.

\section{Command Flow between Front-end and Back-end}
\label{sec:flow}
\begin{figure}[htbp]
\centering
\includegraphics[width=.8\textwidth]{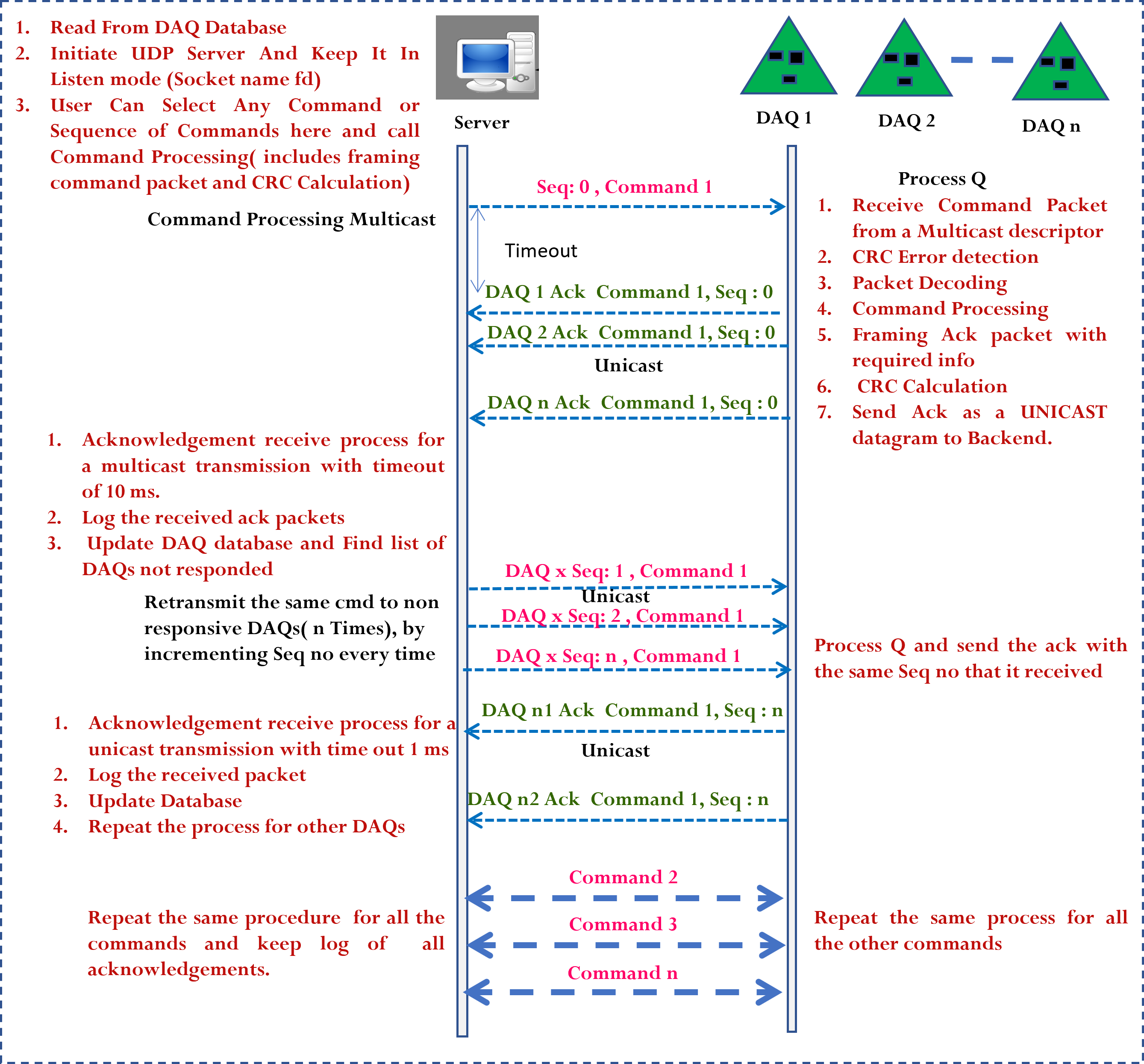}
\caption{Command Transfer between back-end and Frontend.\label{fig:3}}
\end{figure}

HPCI has been implemented as a central command server to oversee the control of all the DAQs. In the context of this paper, we will refer to the central command server as the \textbf{backend} and the RPC-DAQ modules as the \textbf{frontend}. The HPCI command execution follows a sequence of steps in both Front-end and Back-end as illustrated in Figure~\ref{fig:3}. A dedicated UDP server is setup to issue user selected command from the back-end. Receiving these commands the corresponding DAQ carries out series of tasks in the front-end and sends back an acknowledgment. CRC error checking is done for both command and acknowledgment packets. In case of negative or no acknowledgment the command is retransmitted in Unicast mode and the DAQ database is updated with only accessible DAQ list. 

The backend command server maintains a local database that stores the list of currently installed and configured RPC-DAQ modules. This database includes metadata such as DAQ IDs, network addresses, and status flags. It is continuously updated based on command acknowledgments and system responses, and serves as the primary reference for all control operations. For example, when the user needs to find the active status of all the DAQs, this task is accomplished by invoking the \texttt{ISDAQUP} command. The backend server first reads the DAQ list from the local database and sends the \texttt{ISDAQUP} command in multicast mode to all entries. Each DAQ that receives the command responds with acknowledgment packet which consists of DAQ-ID.  After receiving acknowledgments, the command server updates the local DAQ database. The "ISDAQUP" command will be retransmitted in Unicast mode to all DAQs that have not responded within the specified timeout period. During the run start sequence, series of commands are issued to all the DAQ nodes listed in the updated database. The command flow process starts with a group command followed by sequence of Unicast commands.  The payload size varies depending on the command type, as depicted in Figure~\ref{fig:6} and listed in Table~\ref{tab:3}. Some of the UDP command packets don't have payload because they are used only to retrieve DAQ information. The UDP command packet size varies from minimum 18 bytes to maximum 100 bytes. 
While UDP does not guarantee in-order delivery of packets the HPCI protocol incorporates a field named \texttt{Sequence No} in each command and acknowledgment packet to maintain application level ordering and retransmission tracking. The sequence number is incremented by the command server whenever a command is sent or retransmitted. Each DAQ responds with the same sequence number in its acknowledgment packet, allowing the server to:
\begin{itemize}
  \item Match acknowledgments to specific command instances.
  \item Detect duplicate or late arriving acknowledgments.
  \item Ignore stale packets from previous command cycles.
\end{itemize}
This approach effectively maintains in-order execution and consistency of the command state across all DAQs, particularly important during run-start sequences or critical configuration updates. The logic ensures that commands are processed and acknowledged in strict order at the application level, despite using the unordered nature of UDP as the transport protocol.

\section{Implementation and Testing of the HPCI}
\label{sec:Result}
During peak acquisition conditions in the INO ICAL detector each RPC-DAQ module is expected to generate data at a rate of up to 10\,kHz, with typical event packet sizes of 600\,bytes. This results in a per-DAQ data rate of approximately 45\,Mbps. When scaled across several thousand DAQs, the aggregate network load leads to several Gbps, particularly at Level-2 switches. Such high-volume TCP traffic significantly increases the likelihood of congestion, packet buffering delays and retransmission loops especially if TCP is used for command and control alongside event data. To address this, the HPCI protocol is designed to operate independent of the TCP event stream. Its UDP-based architecture minimizes processing overhead, avoids connection setup latency and supports multicast for efficient broadcast command delivery.

\begin{table}[htbp]
\centering
\caption{Cycle time measurement Non-Busy Model.\label{tab:1}}
\smallskip
\begin{tabular}{|c|c|c|}
\hline
Command Type &DAQ Command &Average Cycle time\\
            &Processing Time ($\mu$s) & of a command ($\mu$s) \\
\hline
Command A (18 Bytes) & 174 & 300\\
Command B (26 Bytes) & 405 & 537\\
Command C (100 Bytes) & 615 & 813\\
\hline
\end{tabular}
\end{table}

\begin{table}[htbp]
\centering
\caption{Cycle time measurement Busy Model.\label{tab:2}}
\smallskip
\begin{tabular}{|c|c|c|}
\hline
Command Type &Network Load (Mbps)&Average Cycle time of a command ($\mu$s) \\
\hline
Command C (100 Bytes) & 6kHz * 600Bytes = 27 & 854\\
Command C (100 Bytes) & 10kHz * 600Bytes = 45 & 864\\
\hline
\end{tabular}
\end{table}
The functional validation of the HPCI protocol was performed on a test setup involving five RPC-DAQ modules connected using a gigabit Ethernet switch to a backend server, as depicted in Figure~\ref{fig:2}. The server was equipped with an Intel Xeon processor,16GB RAM running Scientific Linux and an Intel 10G NIC. The RPC-DAQ modules made up of Intel Cyclone IV FPGAs and Wiznet W5300 Ethernet controllers are powered by laboratory DC power supplies.
It is important to note that these measurements were not designed for extensive performance benchmarking. The primary purpose of this test was to validate the behavior of the HPCI command protocol under two scenarios: (i) with minimal network traffic (non-busy mode), and (ii) under synthetic event data load (busy mode). The average command cycle times shown in Tables~\ref{tab:1} and~\ref{tab:2} were observed across multiple runs but standard deviations were not computed, as the focus was strictly on protocol functionality and timing feasibility.

This preliminary testing informed the selection of reasonable timeout thresholds and retransmission logic in the protocol, ensuring reliable operation under expected network loads during full-scale deployment.

\subsection{Command Processing at the Back-end}
\label{sec:cmd_be}
Managing commands in the back-end server of the HPCI model involves  communicating with a large number of DAQs simultaneously. The back-end command server is crucial for receiving acknowledgments from these DAQs and keeping a detailed database to monitor their health status. Also, it 
handles the retransmission of non-acknowledged commands. 

\begin{figure}
    \centering
    \begin{minipage}{0.45\textwidth}
        \centering
        \includegraphics[width=1\textwidth]{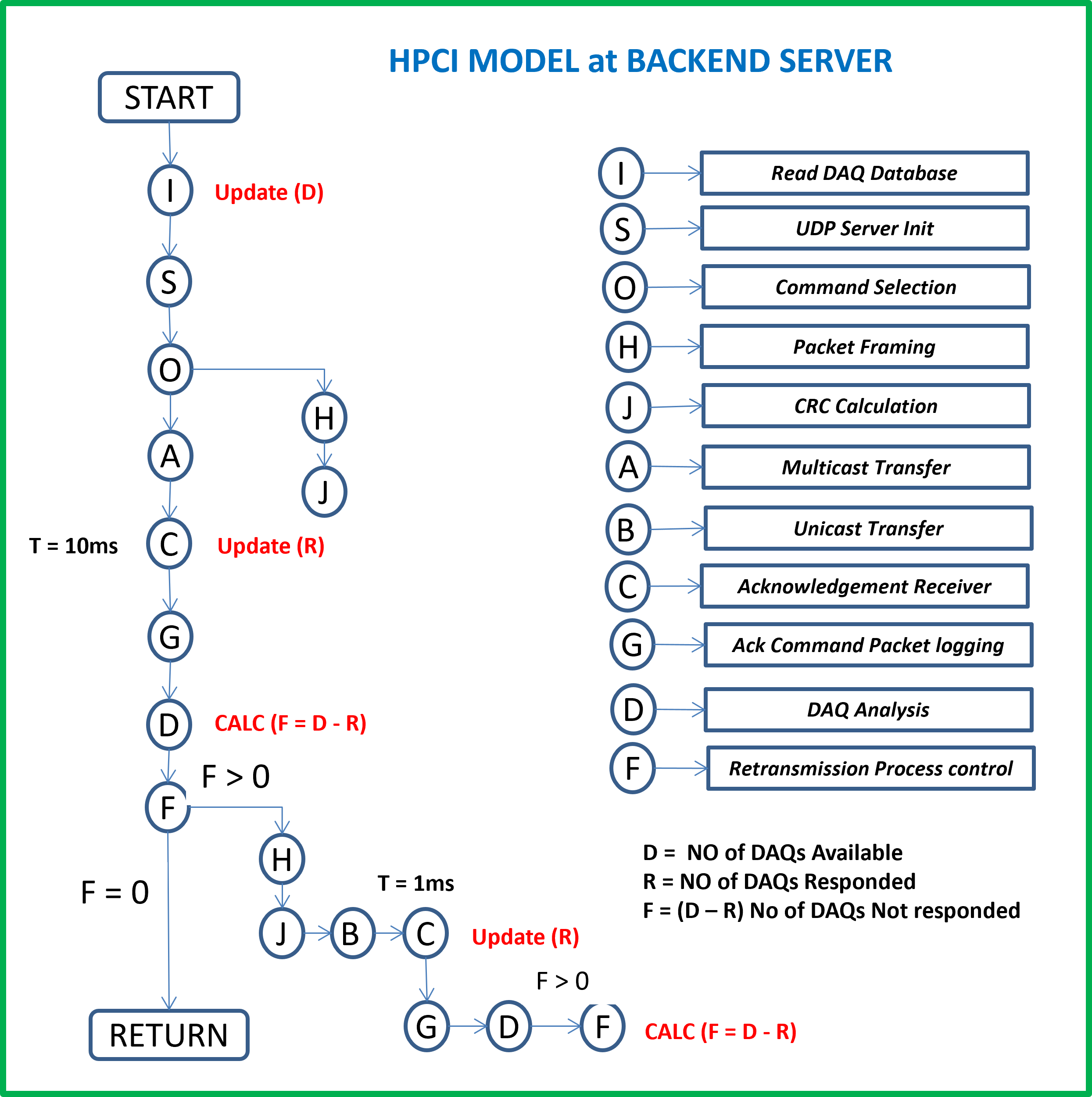} 
        \caption{Command Processing in Back-end.}
        \label{fig:4}
    \end{minipage}\hfill
    \begin{minipage}{0.45\textwidth}
        \centering
        \includegraphics[width=0.95\textwidth]{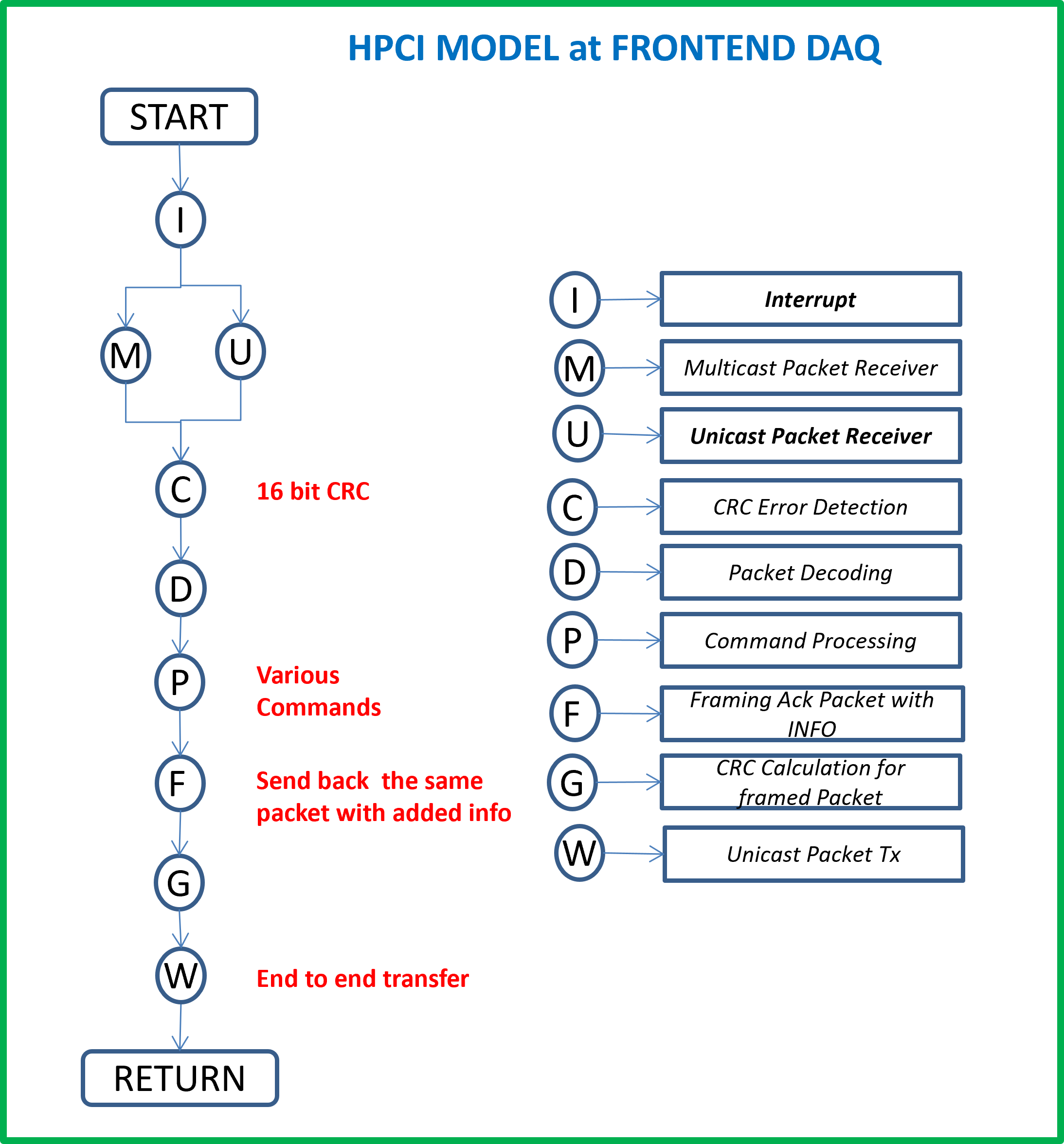} 
        \caption{Command execution in Front-end.}
        \label{fig:5}
    \end{minipage}
\end{figure}

Command processing at the back-end is executed sequentially including steps like reading database, setting up UDP server, selecting payload and adding CRC word. At the start  the run control process consults the pre-loaded DAQ database, extracting information and formulating an expected DAQ list (D).  The back-end server creates a dedicated UDP socket to transmit the command in Multicast mode. The Acknowledgment reception process then captures acknowledgment responses from each DAQ, facilitating the creation of an active DAQ list (R). A comparative analysis of lists D and R enables the back-end to compile a roster of failed DAQs (F). The retransmission process takes charge of each DAQ on the F list, reissuing the same command for a specified number of iterations (n). Finally, a conclusive DAQ database is constructed, based on the presently available DAQs. This database equips users with the information to initiate subsequent commands. This sequential operation, shown in Figure~\ref{fig:4}, allows users to make informed decisions based on real-time DAQ status information.


\subsection{Command Processing at the Front-end DAQ}
\label{sec:cmd_fe}
In the HPCI model, each front-end DAQ is equipped with an identical command processing function. The Frontend RPC-DAQ module comprises of an Intel Cyclone \Romannum{4} FPGA and Wiznet W5300 for Ethernet interface. These frontend DAQs are designed with interrupt capabilities, upon receiving a command the Ethernet controller, Wiznet, generates an interrupt signal to the Softcore processor NIOS instantiated within the FPGA. The Processor Interrupt Service Routine (ISR), illustrated in Figure~\ref{fig:5}, has three major tasks: Command Decoding (D), Command Processing (P) and Acknowledgment Generation (F). Each of these tasks  takes around 100~\(\mu\)s of processing time. However, it's important to ensure that the DAQs can continue performing other essential functions without interruption. To efficiently handle command and DAQ processing a simple handshake scheme is employed in DAQ processor. When a command is received by the DAQ processor, the ISR executes CRC error checking on the received data and sets the "Command arrival flag" to high before exiting the ISR. The remaining processing of command (D,P,F) is executed in the main loop by checking the "Command arrival flag". This flag is set to zero after the command execution in main loop. This method efficiently handles the DAQ processing time by handling every command as well as other high priority tasks.

\subsection{Critical Command Processing and Error Detection}
\label{sec:crc}
\begin{figure}[htbp]
\centering
\includegraphics[width=1\textwidth]{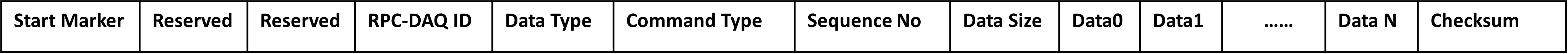}
\caption{Command and Acknowledgment Packet.\label{fig:6}}
\end{figure}

The command interface plays an important role in executing detector specific tasks within the DAQ. Every command packet has a "Command Word" field  as illustrated in Figure~\ref{fig:6} and Table~\ref{tab:3}. Depending on the command word, the RPC-DAQ performs a specific task that alters crucial aspects of data acquisition. While some commands are fundamental for health monitoring, in scenarios where these tasks are exceptionally critical, cannot be duplicated, and involve highly sensitive parameters (where loading incorrect parameters could lead to system damage), the HPCI incorporates command log services and CRC checksum to address these challenges effectively.

\begin{table}[htbp]
\centering
\caption{Command Data Structure.\label{tab:3}}
\smallskip
\begin{tabular}{l|l}
\hline
Command word & Description\\
\hline
Start Marker & Command : 0xDDDD , Acknowledgement : 0xEEEE\\
Reserved & Reserved for later use in ICAL\\
RPC-DAQ ID & DAQ number configured in each RPC-DAQ\\
Data Type & 0xFF00/BB00+0000/00AA (Forward/Backward+ NoACK/Ack )\\
Command Word & Command Identification word for various command\\
Sequence No	& Increments when the command is resent by the Command server\\
Data Size	& Size of the payload Data 0…N\\
Data 0 to N	& Actual message (variable in size depends on the command)\\
Checksum	& CRC-16 Calculated for the whole packet before transmitting on either side\\

\hline
\end{tabular}
\end{table}

In addition to the DAQ database maintained by the backend server, the system also maintains detailed command logs for each RPC-DAQ. These logs serve as a historical record of issued commands, including timestamps, DAQ responses, acknowledgment status, and retry attempts. They do not replace the database used for real-time command execution, but rather support debugging, diagnostics, and traceability during and after data-taking sessions.
Although the Ethernet protocol includes a robust 32-bit CRC mechanism at the frame level, this check is typically performed at the physical and data link layers, and errors detected at this level prevent the packet from reaching the upper layers. In systems using commercial IP cores, such errors are rare due to hardware-level filtering. However, in the HPCI model, additional integrity concerns arise at the application layer, particularly due to the use of embedded FPGA systems with custom UDP implementations and interrupt-driven data paths. 

To address these issues, we include a 16-bit CRC (CRC-16, polynomial 0x8005) in the payload of each command and acknowledgment packet. This check covers the full application-layer packet structure and serves to detect inconsistencies that may arise due to:
\begin{itemize}
  \item Buffer corruption within the NIOS soft-core or Wiznet W5300 interface.
  \item Errors introduced during ISR readout, packet parsing, or DMA transfers in the FPGA.
  \item Data mismatches during retransmission and timeout handling logic.
\end{itemize}

Among the various checksum methods studied such as Fletcher's checksum, longitudinal redundancy check (LRC) and simple additive checks, CRC-16 was selected for its proven balance of error detection capability and computational efficiency in embedded systems~\cite{koopman_crc}. While we have not encountered any corrupted Ethernet frames in the current testbed, the CRC-16 adds an extra layer of validation for critical command control operations and protects against rare but harmful internal faults not caught by the hardware CRC of Ethernet.

\section{HPCI in mini-ICAL}
\label{sec:mical}
mini-ICAL~\cite{mical}, located at the IICHEP Transit Campus in Madurai, Tamil Nadu, India, serves as a prototype for the proposed ICAL. mini-ICAL is a scaled down version, approximately 1/600th the size of ICAL. It consists of 11 layers of iron plates with each layer housing two RPCs and their respective electronics in each slot. One of the objectives of mini-ICAL is to address engineering challenges particularly in the context of data acquisition and detector performance in the stray magnetic field of ICAL.

\begin{figure}[htbp]
\centering
\includegraphics[width=.8\textwidth]{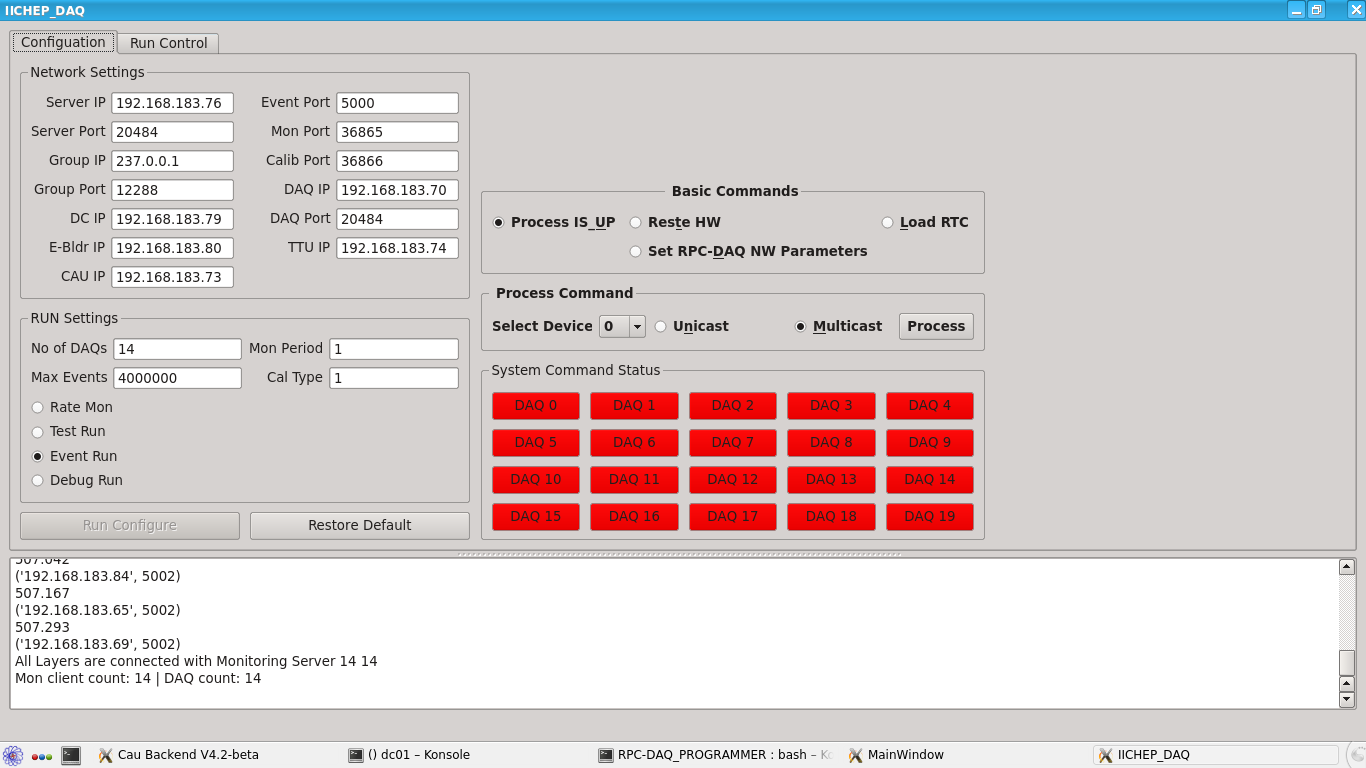}
\caption{Run Control GUI of mini-ICAL Experiment.\label{fig:7}}
\end{figure}

RPC-DAQ is strategically positioned to acquire and process data from the detectors. These modules receive power, Ethernet connectivity, global clock signals and triggers from the Back-end ~\cite{mical}.
To facilitate communication and networking a 24-port 10Gbit Ethernet Switch operates as an L1 Network Switch, establishing a local area network (LAN). Additionally, mini-ICAL integrates various subsystems and servers into this LAN, connecting them via another L2 switch. The heart of mini-ICAL data acquisition system lies in its Run control software which uses HPCI for efficient command control to the RPC-DAQs and subsystems. Figure~\ref{fig:7} shows the Run Control Graphical User Interface (GUI), which provides a user-friendly layout with dedicated tabs for configuration and control. This software controls various components of mini-ICAL by distributing commands and managing command logs, retransmissions and error detection. The run control software comprises two windows or tabs. One window allows for the modification of network and detector configurations, while the other controls the run start stop sequence and display of useful data. Command acknowledgments are highlighted using color coding where green denotes success, and red indicates failure. A run console is included, where the software logs and displays details of each command execution. After its first installation in 2018 the run control software is being used for day by day data taking in mini-ICAL. Throughout this period, several key issues were identified and addressed, particularly adjustments made to acknowledgment timeouts to enhance the software performance.

\section{Timing Precision and Reliability Requirements of INO ICAL Experiment}

The INO ICAL experiment calls for strict requirements on command distribution latency and communication reliability to ensure synchronized operation of its 28,800 RPC-DAQ modules. Commands such as run start, high voltage configuration and synchronization signals must be executed across all nodes with minimal delay and high confidence.

Based on system functionality studies and detector synchronization needs the HPCI command interface is designed to operate with response latencies of less than 1\,ms per DAQ under expected data acquisition loads. Command reliability is further enhanced by incorporating application level CRC-16 error detection, acknowledgment tracking and conditional retransmission. These features ensure a delivery success rate exceeding 99.99\% for critical control commands.

This level of timing precision and reliability supports time-aligned event tagging, coordinated data-taking and robust detector health monitoring across the full DAQ network all of which are essential for the physics goals of the INO ICAL experiment.

\section{Other Responsibilities of Command server in ICAL Experiment}
\label{sec:other}
HPCI allows the command server to configure run conditions including setting up TCP sockets for event acquisition, enabling periodic monitoring, and facilitating remote system upgrades. These processes are essential for seamless data collection and system maintenance. HPCI is instrumental in distributing epoch time to all the DAQs. This timestamp is used by the Real Time Clock (RTC) hardware logic in RPC-DAQ to precisely synchronize commands and events. This helps in accurate event time-stamping and data correlation by the event servers. The Detector parameters like high voltage (HV) and channel masking are configured and monitored through HPCI. This allows the detectors to operate with specified parameters and maintain optimal condition. 
Due to the large number of RPC-DAQ modules in the INO ICAL experiment, a scalable configuration management is needed. Each DAQ is equipped with programmable flash memory that stores critical network parameters including IPv4 address, MAC address, subnet mask, gateway and multicast group information. These settings are read by the embedded soft-core processor during power-up, ensuring the module initializes with correct network settings.
To handle remote management, HPCI Back-end supports a dedicated set of configuration commands that allow the central server to modify these parameters dynamically. This configuration mechanism is integrated within the run-control software, which acts as a centralized utility for managing DAQ identity and network mapping. The DAQ configuration can be updated over the network and reloaded during controlled reboot cycles.
While general-purpose configuration tools like Puppet or Ansible are typically used for server clusters, our approach reflects the practical constraints of embedded systems without operating systems. A similar mechanism is used in FPGA-based systems such as IPBus~\cite{larrea2015ipbus} and the RFU command protocol~\cite{rfu2024tcpip}, where networked control and addressability are managed directly through firmware-level command stacks.

\section{Conclusion}

Although the HPCI approach is not a novel concept  in network-based DAQs, its application within the INO ICAL project presents unique challenges due to the sheer scale of network-enabled DAQs involved. As a consequence, HPCI is devised with traditional command interface approach flavored with smarter handshaking techniques. Currently, HPCI based command server is scaled and tested up to 20 RPC-DAQs in mini-ICAL.  However, the real test lies ahead as it will soon be deployed and evaluated with a significantly larger number of DAQs within the INO ICAL project. The HPCI continues to evolve and be time tested in mini-ICAL.


\acknowledgments
We sincerely thank all our present and former INO colleagues especially Puneet Kanwar Kaur, Umesh L, Anand Lokapure, Aditya Deodhar, Sagar Sonavane, Suraj Kole, Salam Thoi Thoi, Rajkumar Bharathi for technical support during the design, Also we would like to thank members of TIFR namely S.R. Joshi, Piyush Verma, Darshana Gonji, Santosh Chavan and Vishal Asgolkar who supported testing and commissioning. Also we like to thank former INO directors N.K. Mondal and V. M Datar for their continuous encouragement and guidance. 
.  


\end{document}